\documentclass[12pt,a4paper]{article}
\pdfoutput=1
\usepackage{jheppub}

\usepackage{textcomp}
\usepackage{amsmath, amssymb}
\usepackage{bbold}
\usepackage{mathbbol}
\usepackage{slashed}
\usepackage{graphics}
\usepackage{shadow}
\usepackage{float}
\usepackage{latexsym,feynmf}
\def\sideremark#1{\ifvmode\leavevmode\fi\vadjust{\vbox to0pt{\vss
 \hbox to 0pt{\hskip\hsize\hskip1em
 \vbox{\hsize3cm\tiny\raggedright\pretolerance10000
 \noindent #1\hfill}\hss}\vbox to8pt{\vfil}\vss}}}%

\newcommand{\comment}[1]{}

\newcommand{\I}{\textbf{I}}
\newcommand{\It}{\widetilde{\textbf{I}}}
\newcommand{\de}{\partial}
\newcommand{\be}{\begin{equation}}
\newcommand{\ee}{\end{equation}}
\newcommand{\bea}{\begin{eqnarray}}
\newcommand{\eea}{\end{eqnarray}}

\def\ccr{\nonumber\\}

\title{Effective action for higher spin fields on (A)dS backgrounds}
\author[a]{Fiorenzo Bastianelli,}
\author[a]{Roberto Bonezzi,}
\author[b,c]{Olindo Corradini,}
\author[d,e]{Emanuele Latini}

\affiliation[a] {Dipartimento  di Fisica, Universit{\`a} di Bologna and\\
INFN, Sezione di Bologna, via Irnerio 46, I-40126 Bologna, Italy}
\affiliation[b] {Centro de Estudios en F\'isica y Matem\'aticas Basicas y Aplicadas\\
Universidad Aut\'onoma de Chiapas, Tuxtla Guti\'errez 29000, Mexico}
\affiliation[c] {Dipartimento di Fisica, Universit\`a di Modena\\ Via Campi 213/A, I-41125 Modena, Italy}
\affiliation[d] {Institut f{\"u}r Mathematik, Universit{\"a}t Z{\"u}rich-Irchel, Winterthurerstrasse 190, CH-8057 Z{\"u}rich, Switzerland}
\affiliation[e] {INFN, Laboratori Nazionali di Frascati, CP 13,
I-00044 Frascati, Italy}

\emailAdd{bastianelli@bo.infn.it}\emailAdd{bonezzi@bo.infn.it}\emailAdd{olindo.corradini@unach.mx}
\emailAdd{emanuele.latini@math.uzh.ch}

\abstract{We study the one loop effective action for a class of higher spin fields by using a
first-quantized description. The latter is obtained by considering spinning particles,
characterized by an extended local supersymmetry on the worldline, that can propagate consistently
on conformally flat spaces. The gauge fixing procedure for calculating the worldline path integral on a loop
is delicate, as the gauge algebra contains nontrivial structure functions.
Restricting the analysis  on (A)dS backgrounds simplifies the gauge fixing procedure, and allows us to produce a useful
representation of the one loop effective action. In particular, we extract the first few heat kernel coefficients for arbitrary even spacetime dimension $D$ 
and for spin $S$ identified by a curvature tensor with the symmetries of a rectangular Young tableau of $D/2$ rows and $[S]$ columns.}

\keywords{Sigma Models, Extended Supersymmetry, Field Theories in Higher Dimensions}

\begin{document}
\maketitle
\flushbottom

\section{Introduction}
\label{sec:intro}
Higher spin field theory is a topic that enters several aspects of modern theoretical physics. In this paper we quantize higher spin fields on (A)dS spaces using a worldline approach and study their one loop effective action,  extending the analysis of \cite{Bastianelli:2007pv} that was restricted to flat spacetimes.

The worldline approach to quantum field theory 
(see~\cite{Schubert:2001he} for a review), has been known to be an
alternative tool to compute Feynman diagrams through the quantization
of relativistic point particles. More specifically,
one loop effective actions in the presence of external fields find an
efficient approach in terms of point particle path integrals computed
on the circle, whereas field theory propagators are linked to particle
path integrals on the line. 
 In particular, for relativistic higher
spin fields (see~\cite{Vasiliev:2004qz,Sorokin:2004ie,Bouatta:2004kk,Bekaert:2005vh,Fotopoulos:2008ka,Sagnotti:2011qp}
 for reviews) the particle approach might be particularly useful to
extract information beyond the classical level.  It is the main objective of the present manuscript to
use a particle approach to compute the one loop effective action for
higher spin fields in curved space.
Indeed, extensions of the worldline approach to field theories with background gravity are feasible, as discussed for example 
in~\cite{Bastianelli:2002fv, Bastianelli:2002qw, Bastianelli:2003bg, Bastianelli:2005vk, Bastianelli:2005uy, Bastianelli:2004zp}.

The class of higher spin particles that we wish to treat here are those  described by the 
 $O(N)$ spinning particles actions~\cite{Berezin:1976eg, Gershun:1979fb, Howe:1988ft, Howe:1989vn},
that contain a fully-gauged extended supersymmetry on the worldline.
These models describe in first quantization higher spin fields that enjoy conformal invariance in flat spacetimes 
\cite{ Siegel:1988ru, Siegel:1988gd, Metsaev:1995jp}. They form the complete set in $D=4$, and for
spin $S>1$ they live only in even space-time dimensions.
In \cite{Bastianelli:2008nm} the conformal invariance was proven by showing that these particle models have classical background reparametrization 
and Weyl invariance, thus leaving the conformal Killing vectors as generators of true symmetries. This result also implies that
 these models are consistent on generic conformally flat spaces. The particular coupling to (A)dS spaces  was previously known from the work of
 \cite{Kuzenko:1995mg}. The class of higher spin fields treated here 
 can be described by higher spin curvature tensors that obey the symmetries of a Young tableau of $D/2$ rows and $[S]$ columns
 (see \cite{Corradini:2010ia}  for a discussion of the curvature tensors for half-integer spin). 
More general types of higher spin fields could perhaps be described 
by using the detour worldline methods of~\cite{Bastianelli:2009eh,Cherney:2009mf,Cherney:2009md}.  
 
The gauge structure of our particle models on generic conformally flat spaces is quite complex, as it contains non-trivial structure 
functions  \cite{Bastianelli:2008nm}.
We find it  simpler, for the moment being, to  investigate the one loop effective action on maximally symmetric spaces, i.e. (A)dS spaces, which allow for an
algebraically simpler gauge fixing procedure. Weyl anomalies are generically present in quantum field theories, so that we expect to find 
a nontrivial  effective action, as indeed we do. 

One may also approach the problem directly in quantum field theory, as suitable actions are known, see for example 
\cite{Fronsdal:1978rb, Fang:1978wz, Fronsdal:1978vb, Buchbinder:2001bs, Francia:2005bu, Buchbinder:2007vq, Campoleoni:2008jq, Campoleoni:2009gs, Campoleoni:2012th}.
 However we wish to suggest here the point of view that  many results are more efficiently obtained using first quantized methods.\footnote{
 A worldline approach to quantum massive higher spins in (A)dS ~\cite{Deser:2001pe, Deser:2001us, Zinoviev:2001dt,de Medeiros:2003px,Metsaev:2003cu}
can be treated along similar lines by dimensionally reducing the O(N) spinning particle used here.}
Recently the heat kernel for some higher spin fields  in (mostly) odd-dimensional maximally symmetric spaces were computed using a group-theoretical approach~\cite{David:2009xg,Gopakumar:2011qs,Gupta:2012he}. Our approach deals with a different set of multiplets on even-dimensional spaces. It would be useful to eventually compare the two approaches. Also, a different type of effective action with higher spin backgrounds was studied in~\cite{Bekaert:2010ky}.
 
In subsequent sections we first
present the gauge fixing of the models under study, then briefly
review the regularization techniques needed to compute worldline path integrals in curved spaces. 
Finally we present the derivation of the worldline representation of the effective action.
It is generically difficult to compute it in a closed form, so
we aim here to calculate explicitly only the first few heat kernel coefficients for (A)dS backgrounds.
For $D>2$ these correspond to diverging terms that must be subtracted to renormalize the effective action.
We perform the path integral computation with an arbitrary metric, as intermediate calculations might be useful
for a larger class of backgrounds. Indeed, as mentioned above, these spinning particles are certainly consistent on conformally flat spaces.
However, in that case the gauge fixing procedure is much more laborious and will not be attempted here.

The present analysis could be repeated step by step to carry out similar 
calculations for the $U(N)$ spinning particle \cite{Marcus:1994mm},
which gives rise to higher spin fields living on complex spaces
\cite{Bastianelli:2009vj} (treated already for the particular cases of N=1,2 
on arbitrary Kahler manifolds in \cite{Bastianelli:2011pe,Bastianelli:2012nh}). 

To conclude, the main results derived here are a worldline
representation of the one loop effective action 
for a class of higher spin fields on (A)dS spaces, see eq.~\eqref{eq:nlsm},
and the calculations of the first few heat kernel coefficients, 
see eqs.~(\ref{sdweven},\ref{sdweven1}) for integer spin and eqs.~(\ref{sdwodd},\ref{sdwodd1}) for half-integer spin.

\section{Spinning particle on conformally flat spaces}
\label{sec:review}

The model we study here is the (fully) gauged
counterpart of the mechanical model with action
\begin{eqnarray}
  S &=& \int_0^1 dt \Big( p_\mu \dot x^\mu  +\frac i2
  \psi_i^a \dot\psi_{ia} -\frac12 p_\mu p^\mu \Big)~,\quad i=1,\dots,N
\label{eq:action-sf}
\end{eqnarray}
where a set of global worldline symmetries (time translation, $N$
supersymmetries, $O(N)$ $R$-symmetry) are rendered local to guarantee unitarity.
Here $x^\mu$ and $p_\mu$ are spacetime coordinates
and momenta of the particle, whereas $\psi_i^a$ are Majorana
fermions, with $a$ a flat Lorentz index. The resulting phase-space action identifies the so-called $O(N)$ spinning particle model and, when considering a curved target space, reads
\begin{eqnarray}
  S[x,p,\psi,E;g] &=& \int_0^1 dt \Biggl[ p_\mu \dot x^\mu  +\frac i2
  \psi_i^a \dot\psi_{ia}-eH-i \chi_i\underbrace{\pi_\mu e_a^\mu\psi_i^a }_{Q_i} -\frac12
    a_{ij}\underbrace{i\psi_i\cdot\psi_j}_{J_{ij}}\Biggr]
  \label{eq:action-ps}
\end{eqnarray}
with $H=H_0 -\frac18 R_{abcd} ~\psi^a \cdot \psi^b \psi^c \cdot
\psi^d$ and $H_0=\frac12 g^{\mu\nu}\pi_\mu \pi_\nu$ being the kinetic hamiltonian written in terms of the covariant  momenta $\pi_\mu = p_\mu -\frac{i}{2}\omega_{\mu
  ab} \psi^a_i\psi^b_i$. From~\eqref{eq:action-ps} one recognizes the supercharges $Q_i$ and the $O(N)$ symmetry generators $J_{ij}$.
   $E$ collectively denotes the worldline gauge fields $E=(e,\chi_i,a_{ij})$,
i.e. einbein, gravitini and $O(N)$ gauge fields respectively. This
model describes the first quantization of a particular mixed-symmetry
higher spin particle in $D=2d$ even-dimensional curved space, that generically (for $N > 2$) must be conformally flat. The spectrum of the model for $N>2$ is empty in odd dimensions~\cite{Howe:1989vn}. For even $N=2n$ the model describes equations of motion (the Dirac constraints) for a bosonic field strength characterized by a rectangular Young tableau with $n$ columns and $d$
rows. For odd $N=2n+1$ the model describes equations of motion for a fermionic field strength, a spinor-tensor with a tensor structure characterized by the same $n\times d$ Young tableau.  For $D=4$ this
involves all possible massless representations of the Poincar\'e group, that at the level of gauge potentials are given by 
totally symmetric (spinor-) tensors, whereas for $D>4$ it corresponds to
conformal multiplets only~\cite{Siegel:1988ru,Siegel:1988gd,Metsaev:1995jp}. The euclidean configuration
space action, that one obtains after integrating out the momenta
$p_\mu$ and Wick rotating, reads
\begin{eqnarray}
  S[y,E;g] &=& \int_0^1 d\tau \Biggl[ \frac1{2e} g_{\mu\nu}
  \Big( \dot x^\mu-\chi_i\psi^\mu_i \Big) \Big( \dot
    x^\nu-\chi_j\psi^\nu_j \Big)  \label{eq:action-cs}
\\ &&\hskip1cm+\frac12
  \psi_i^a\Big(\delta_{ij}\delta_{ab}\partial_\tau
    +\dot x^\mu \omega_{\mu ab} \delta_{ij} -a_{ij}\delta_{ab}
  \Big)\psi_j^b 
 -\frac e8 R_{abcd}~\psi^a\cdot
  \psi^b \psi^c\cdot\psi^d\Biggr]
\nonumber
 \end{eqnarray}
 with $y=(x^\mu,\psi_i^a)$ being the ``matter" fields. For arbitrary
 $N$ and generic curved backgrounds the gauge symmetry generators
 $(H,Q_i,J_{ij})$ do not form a first class algebra. However in~\cite{Bastianelli:2008nm} it was found
 that, if the background is conformally flat, they
 form a (nonlinear) first-class constraint algebra and the previous
 action is gauge-invariant under the transformations induced by the
 gauge symmetry generator $G=\xi H +i\epsilon_i Q_i +\frac12
 \alpha_{ij} J_{ij} := \Xi^A G_A $.\footnote{For $N\leqslant 2$
   the $R$-symmetry group is either trivial or abelian, and the algebra
   closes on an arbitrary background.} 
   
   At the quantum level the
constraint algebra on conformally flat spaces closes as well, provided
one adds to the hamiltonian an improvement term proportional to the
scalar of curvature, namely
\begin{eqnarray}
  H=H_0 -\frac18 R_{abcd} ~\psi^a \cdot \psi^b \psi^c \cdot \psi^d-
   \frac{(N-2)(D+N-2)}{16(D-1)} R
\label{eq:qH}
\end{eqnarray}
with the kinetic operator given by
\begin{eqnarray}
  H_0 &=&\frac12\Big(\pi^a-i\omega_b{}^{ba}\Big)\pi_a \nonumber\\
  \pi_a &=& e_a^\mu \pi_\mu \;, \quad \pi_\mu = g^{1/4}
  p_\mu g^{-1/4}
  -\frac{i}{2}\omega_{\mu ab} \psi^a_i\psi^b_i~.
\label{eq:H0}
\end{eqnarray}

Here we use a path
integral formalism and find it more convenient to use the
(euclidean) configuration space action
\begin{eqnarray}
  S[y,E;g] &=& \int_0^1 d\tau \Biggl[ \frac1{2e} g_{\mu\nu}
    \Big( \dot x^\mu-\chi_i\psi^\mu_i \Big) \Big( \dot
    x^\nu-\chi_j\psi^\nu_j \Big)\nonumber\\&&\hskip1cm +\frac12
    \psi_i^a\Big(\delta_{ij}\delta_{ab}\partial_\tau
    +\dot x^\mu \omega_{\mu ab} \delta_{ij} -a_{ij}\delta_{ab}
    \Big)\psi_j^b -\frac e8 R_{abcd}~\psi^a\cdot
    \psi^b \psi^c\cdot\psi^d\nonumber\\
    &&\hskip1cm- e \frac{(N-2)(D+N-2)}{16(D-1)} R\Biggr]
  \label{eq:action-cs'}
\end{eqnarray}
that is~\eqref{eq:action-cs} with the addition of the improvement
term. The associated path integral evaluated on the circle $S^1$
\begin{eqnarray}
  \Gamma[g] = \int_{_{S^1}} \frac{DE\, Dy}{\rm Vol\, (Gauge)}\ e^{-S[y,E;g] }
  \label{eq:eff-action}
\end{eqnarray}
gives a representation of the one loop effective
action for the aforementioned higher spin field coupled to external
gravity. It is defined by taking bosonic
fields with periodic boundary conditions and
fermionic fields  with antiperiodic boundary
conditions. 

In order to be able to perform computations  two preliminary issues have to be taken care of:  \\[2mm]
i) Firstly, the worldline action must be suitably gauge-fixed;
i.e. the gauge fields $E$ must be fixed to some specific
configuration that will depend upon a set of modular parameters that must
be integrated over. In the
present case the gauge symmetry algebra, associated to the above
generators, is nonlinear, i.e. commutators of pairs of generators
involve structure {\it functions} and not structure {\it
  constants}. Therefore one must use more powerful hamiltonian BRST methods 
  to gauge fix the action in its hamiltonian form. \\
ii) The resulting gauge-fixed action depends only upon ``matter" fields and
modular parameters. However, in curved space, it still is a nonlinear sigma
model, so that for perturbative computations  one usually Taylor expands the metric about a fixed point
of the circle. This results in an infinite set of vertices. In addition some
Feynman diagrams present ambiguities  and need to be
regularized. This is a well-known fact, and several regularization
schemes have been used in the past to compute such path
integrals; see~\cite{Bastianelli:2006rx} for an overall review. Up to recently only quantum-mechanical path integrals in curved space with $N\leqslant 2$ had been
used: these path integrals, in the worldline  formalism, correspond to
the first quantization of spin $S\leqslant 1$ fields in curved
space. More recently in~\cite{Bastianelli:2011cc} the
regularization of nonlinear sigma models with arbitrary $N$ was considered, having in mind applications to the $O(N)$ spinning particles. 
What studied in~\cite{Bastianelli:2011cc} are the globally supersymmetric counterparts of the
models studied here. That is enough for the present purposes as the gauging does not introduce additional ambiguities.

\subsection{Gauge-fixing in (A)dS}
\label{sec:gfMS}
In this section we describe the gauge-fixing of the $O(N)$ spinning particle propagating on (A)dS spaces.  For such backgrounds the Riemann curvature can be written as
\begin{eqnarray}
  \label{eq:MS}
  R_{abcd} =b (\eta_{ac}\eta_{bd}-\eta_{ad}\eta_{bc})
\end{eqnarray}
where $\Lambda = (D-1)(D-2) b$ is the cosmological constant. Let us start considering the action in hamiltonian form. At the classical level (cfr.~\eqref{eq:action-ps}), in (A)dS spaces the hamiltonian constraint reduces to $H=H_0 -\frac b4
J_{ij}J_{ij}$ and the first-class algebra reduces to a quadratic algebra (curly
brackets here are graded Poisson brackets)
\begin{align}
  \{Q_i,Q_j\} &= -2i \delta_{ij} H +ib \Big (J_{ik}J_{jk}
  -\frac{1}{2}\delta_{ij} J_{kl} J_{kl}\Big)
\nonumber  \\[1mm]
  \{J_{ij},J_{kl}\} &= \delta_{jk} J_{il} - \delta_{ik} J_{jl}
  - \delta_{jl} J_{ik} + \delta_{il} J_{jk}\nonumber \\[1mm] 
  \{J_{ij},Q_k\} &= \delta_{jk} Q_i -\delta_{ik} Q_j\,,\quad  \{H,Q_i\} = \{H,J_{ij}\} =0
  \label{eq:nonlinear-JJ}
\end{align}
that can be used to obtain the corresponding transformations of the gauge
fields.

Upon canonical quantization the latter quadratic algebra turns into the  following (anti-)commutation relations
\begin{align}
\{Q_i,Q_j\} &= 2 \delta_{ij} H
-\frac{b}{2} (J_{ik}J_{jk}+J_{jk}J_{ik} -\delta_{ij} J_{kl}J_{kl}) \ccr[1mm]
[J_{ij},J_{kl}] &= i\delta_{jk} J_{il} - i\delta_{ik} J_{jl}
- i\delta_{jl} J_{ik} + i\delta_{il} J_{jk} \ccr[1mm]
[J_{ij},Q_k]&= i\delta_{jk} Q_i -i \delta_{ik} Q_j\,,\quad 
[H,Q_i] = [H,J_{ij}] =0
\label{eq:quantum-algebra}
\end{align}
with the hamiltonian constraint given by~\eqref{eq:qH}, that in (A)dS reduces to
\begin{eqnarray}
H=H_0 -\frac b4 J_{ij}J_{ij} -bA(D)
\end{eqnarray}
with $A(D)=-\frac{D}{8}(D+N-2)$.

In order to gauge fix the locally symmetric $O(N)$ spinning particle action  (with quantum gauge algebra given in~\eqref{eq:quantum-algebra}) we use the hamiltonian BRST method reviewed in Appendix~\ref{app:brst}. Basically, we
define ghost fields ${\cal C}^A = ({\cal C}, {\cal C}_i,{\cal C}_{ij})$ and ghost
momenta ${\cal P}_A = ({\cal P},{\cal P}_i,{\cal P}_{ij})$
for all constraint generators $G_A = (H,Q_i,J_{ij})$, such that $[{\mathcal P}_A,{\cal C}^B\}=-i\delta_A^B$ and write the quantum BRST operator as a graded sum $\Omega =
\sum_{p\geq 0} \stackrel{_{(p)}}{\Omega}$. Starting from 
\begin{eqnarray}
  \stackrel{_{(0)}}{\Omega} ={\cal C}^A G_A = {\cal C} H + {\cal C}_i Q_i
  +{\cal C}_{ij} J_{ij}
\end{eqnarray}
and imposing the nilpotency of the BRST charge, we can recursively obtain higher antighost-number operators. Setting
\begin{eqnarray}
[G_A, G_B \}=F^C_{AB}(z)\ G_C
\label{eq:quantum-algebra-sf}
\end{eqnarray}
with $z^\alpha=(p_\mu,x^\mu,\psi_i^a)$ and $F^C_{AB}(z)$ structure functions, for the algebra~\eqref{eq:quantum-algebra} we get
 \begin{eqnarray}
  \stackrel{_{(1)}}{\Omega} &=&\frac{i}{2} (-)^{\varepsilon_A}{\cal C}^A
  {\cal C}^B F_{BA}^C {\cal P}_C \nonumber\\ &=& -i{\cal C}_i {\cal C}_i
  {\cal P}-2 {\cal C}_{k} {\cal C}_{ki} {\cal P}_i+2 {\cal C}_{ik}{\cal C}_{kj} {\cal P}_{ij}  -i\frac{b}{4} \Big({\cal C}_i {\cal C}_i J_{kl} {\cal P}_{kl}-2{\cal C}_i {\cal C}_j J_{ik} {\cal P}_{jk}\Big)
\end{eqnarray}
and 
\begin{eqnarray}
\stackrel{_{(3)}}\Omega &=&\frac{b^2}{24}\Big({\cal C}_i {\cal C}_j {\cal C}_k {\cal C}_l\, \mathcal{P}_{ij}\mathcal{P}_{km}\mathcal{P}_{lm}-3{\cal C}_m {\cal C}_m {\cal C}_i {\cal C}_j\,\mathcal{P}_{ik}\mathcal{P}_{jl}\mathcal{P}_{kl}+{\cal C}_k {\cal C}_k {\cal C}_l {\cal C}_l\, Tr(\mathcal{P}_{ij}^3)\Big)\\
\stackrel{_{(2)}}\Omega &=& \stackrel{_{(p)}}\Omega =0\,,\quad p>3~.
\end{eqnarray}
One can thus write the quantum gauge-fixed hamiltonian operator as
\begin{eqnarray*}
{\mathbb H}_{qu} = H_{BRST} -i\{K,\Omega\}
\end{eqnarray*}
where the first term is a BRST-invariant   hamiltonian and $K$ a gauge fixing fermion: the latter is BRST-invariant  for any choice of $K$ thanks to the nilpotency of $\Omega$. In the present case since $H$ itself enters as a constraint we can set $H_{BRST} =0$ and thus have
\begin{eqnarray}
{\mathbb H}_{qu} = -i\{K,\Omega\}~.
\end{eqnarray}
Let us now use the gauge-fixing fermion
\begin{eqnarray}
K=-\hat E^A {\mathcal P}_A\,,\quad \hat E^A=\Big(\beta,0,\frac{\theta_{ij}}{2}\Big)
\end{eqnarray}
with $\theta_{ij}$ a $N\times N$ skew diagonal matrix, dependent on $[S]=[N/2]:= n$ angular variables $\theta_k$, with $k=1,\dots, n$.  Here $S=N/2$ is the ``spin" of the particle. With this choice one obtains the hamiltonian operator
\begin{eqnarray}
{\mathbb H}_{qu} = \beta H +\frac{1}{2}\theta_{ij} J_{ij} - \theta_{ij}{\cal C}_i {\mathcal P}_j-2\theta_{ij} {\cal C}_{im} {\mathcal P}_{jm}
\label{eq:Hqu}
\end{eqnarray}
and consequently the gauge-fixed path integral can be written as
\begin{eqnarray}
\Gamma[g]=K_N \int_0^\infty \frac{d\beta}{\beta}\prod_{k=1}^{n}\int_0^{2\pi}\frac{d\theta_k}{2\pi}\int_{_{S^1}} Dz\, D{\cal C}\, D{\mathcal P}~ e^{iS_{qu}[z,{\cal C},{\mathcal{P}},\hat E;g]}
\end{eqnarray}
with  phase space action
\begin{eqnarray}
S_{qu}[z,{\cal C},{\mathcal{P},\hat E;g}] &=& \int_0^1dt
\Big[ p_\mu \dot x^\mu +\frac{i}{2}\psi^a_i\dot \psi_{ia} +\dot {\cal C}^A {\mathcal P}_A -H_{qu}\Big]\\
H_{qu} &=& \beta \biggl(\frac{1}{2}g_{\mu\nu}(x)\pi^\mu \pi^\nu -\frac{b}{4}J_{ij} J_{ij} -bA(D)\biggr)\nonumber\\&&
+\frac{1}{2}\theta_{ij} J_{ij} - \theta_{ij}{\cal C}_i {\mathcal P}_j-2\theta_{ij} {\cal C}_{im} {\mathcal P}_{jm}
\label{eq:Hcl}
\end{eqnarray}
and $\pi^\mu = p^\mu -\frac{i}{2}\omega^\mu{}_{ab}\psi^a_i \psi^b_i$. Above $K_N$ is a normalization factor that implements the reduction to a fundamental region of moduli space
\begin{align}
K_N=\left\{
\begin{array}{ll}
\frac{2}{2^n n!}\,,\quad & N=2n\\
\frac{1}{2^n n!}\,,\quad & N=2n+1
\end{array}\right.
\end{align} 
as discussed in~\cite{Bastianelli:2007pv}. Integrating out particle momenta leads to a configuration space path integral that involves the action
\begin{eqnarray}
S_{qu}[y,{\cal C},{\mathcal{P},\hat E;g}] &=& \int_0^1dt
\Big[ \frac{1}{2\beta} g_{\mu\nu} \dot x^\mu \dot x^\nu
  +\frac{i}{2} \psi_{a i} D_t \psi^a_i+\beta\Big(\frac{b}{4}J_{ij} J_{ij}+bA(D)\Big)
  -\frac{1}{2}\theta_{ij}J_{ij}\nonumber
\\[1mm]&& -{\mathcal P}\dot {\cal C}+{\mathcal P}_i (\delta_{ij}\partial_t-\theta_{ij}) {\cal C}_j-{\mathcal P}_{ij}(\delta_{im}\delta_{jp}\partial_t-\theta_{im}\delta_{jp}+\theta_{jm}\delta_{ip}){\cal C}_{mp}\Big]
\nonumber\\
\end{eqnarray}
where $D_t \psi^a_i = \dot \psi^a_i
+ \dot x^\mu \omega_{\mu}{}^a{}_b \psi^b_i $.
A Wick rotation to euclidean time yields
\begin{eqnarray}
\Gamma[g]=K_N\int_0^\infty \frac{d\beta}{\beta}\prod_{k=1}^{n}\int_0^{2\pi}\frac{d\theta_k}{2\pi}\int_{_{S^1}} Dy\, D{\cal C}\, D{\mathcal P}~ e^{-S_{qu}[y,{\cal C},{\mathcal{P}},\hat E;g]}
\end{eqnarray}
with the euclidean version of the action given by
\begin{eqnarray}
S_{qu}[y,{\cal C},{\mathcal{P},\hat E;g}] &=& \frac1\beta\int_0^1d\tau
\Big[ \frac{1}{2} g_{\mu\nu} \dot x^\mu \dot x^\nu
  +\frac{1}{2} \psi_{a i} \Big( \delta_{ij }D_\tau -\theta_{ij}\Big)\psi^a_j-\frac{b}{4}J_{ij} J_{ij}-\beta^2 bA(D)
 \nonumber
\\[1mm]&& -{\mathcal P}\dot {\cal C}+{\mathcal P}_i (\delta_{ij}\partial_t-\theta_{ij}) {\cal C}_j+{\mathcal P}_{ij}(\delta_{im}\delta_{jp}\partial_t-\theta_{im}\delta_{jp}+\theta_{jm}\delta_{ip}){\cal C}_{mp}\Big]~.\nonumber\\
\end{eqnarray}
where we have Wick-rotated the $O(N)$ fields $\theta_{ij}\to i\theta_{ij}$ and the ghost momenta ${\mathcal P}_A\to i {\mathcal P}_A$. Here $D_\tau$ is represented by the same covariant derivative as given above, with ``dot" now representing derivative with respect to $\tau$. Fermions and ghosts have been suitably rescaled in order to have a common $\frac1\beta$ in front of the action. In the following we perturbatively compute the above path integral. Although the latter is defined on (A)dS spaces, for convenience we keep the geometry arbitrary and only at the end do we fix it to (A)dS.  In essence, we replace $\frac{b}{4}J_{ij} J_{ij}+\beta^2 bA(D)$ by $\frac18 R_{abcd} \psi^a\cdot \psi^b \psi^c\cdot\psi^d+\beta^2 \frac{(N-2)(D+N-2)}{16(D-1)}R$ in the above action. Integrating over the ghost fields yields
\begin{align}
\Gamma[g]&=K_N \int_0^\infty \frac{d\beta}{\beta}\prod_{k=1}^{n}\int_0^{2\pi}\frac{d\theta_k}{2\pi} \Big({\rm Det}(\partial_\tau -\theta_{\rm vec})_{ABC} \Big)^{-1}{\rm Det'}(\partial_\tau -\theta_{\rm adj})_{PBC}\nonumber\\ &\int_{_{S^1}} {\cal D}x D\psi ~{\rm exp}\Biggl( -\frac{1}{\beta} \int_0^1d\tau
\Big[ \frac{1}{2} g_{\mu\nu} \dot x^\mu \dot x^\nu
  +\frac{1}{2} \psi_{a i} \Big( \delta_{ij }D_\tau -\theta_{ij}\Big)\psi^a_j\nonumber\\
  &-\frac18 R_{abcd} \psi^a\cdot \psi^b \psi^c\cdot\psi^d-\beta^2 \frac{(N-2)(D+N-2)}{16(D-1)}R\Big]\Biggr)
  \label{eq:nlsm}
\end{align}
where $\theta_{\rm vec}$ and $\theta_{\rm adj}$ denote the gauge-fixed $O(N)$ potentials in the vector and adjoint representation, respectively. ${\rm Det'}$ indicates a determinant without its zero modes, and $ {\cal D}x$ is the reparametrization invariant measure. 
Below we consider a short-time perturbative approach to the above nonlinear sigma model path integral.

\subsection{Regularization of supersymmetric nonlinear sigma models}
\label{sec:reg}
For a particle in curved space, the passage between the operatorial representation of the transition amplitude and its path integral counterpart is in general not straightforward, as the latter involves a nonlinear sigma model that perturbatively gives rise to superficial divergences.
These divergences are compensated by vertices arising from the nontrivial path integral measure, but finite ambiguities remain that need to be 
dealt with by specifying a regularization scheme.  This is well studied for models with global (super)symmetries (see~\cite{Bastianelli:2006rx} for a review).
 However it is clear that gauging does not introduce further divergences. 
 Indeed upon gauge fixing,  the gauged model reduces essentially to the ungauged one. 
 Moreover the ghosts do not couple to the target space geometry and just produce the correct measure for integration over the moduli space.
 
 In~\cite{Bastianelli:2011cc} we considered the regularization of the spinning particle
model with hamiltonian
\begin{equation}
  H=H_0+\alpha R_{abcd} \psi^a_i\psi^b_i \psi^c_j\psi^d_j +V
  \label{eq:H}
\end{equation}
with $H_0$ given by~\eqref{eq:H0}. The corresponding euclidean classical action in configuration space is given by
\begin{eqnarray} \label{action1.3}
  && S=  \frac1\beta\int^1_0 d\tau\, \Biggl [
  \frac12 g_{\mu\nu} \dot x^\mu \dot x^\nu
  +\frac{1}{2} \psi_{a i} D_\tau \psi^a_i   +\alpha
  R_{abcd}\psi^a_i\psi^b_i\psi^c_j\psi^d_j +\beta^2 V \Biggr ]
\end{eqnarray}
and, for $\alpha = -\frac18$, is nothing but the ungauged version of the nonlinear sigma model  of the previous section.
 We found that such path integral reproduces the transition amplitudes that satisfies the
Schr\"odinger equation with hamiltonian~\eqref{eq:H} provided we add the
counterterm
\begin{eqnarray}\label{eq:VCT}
  V_{CT} =
  \left\{
    \begin{array}{ll}
      -\left(\frac18+\frac{\alpha N}{2} \right)R+\frac18 g^{\mu\nu}
      \Gamma_{\mu\lambda}^\rho\Gamma_{\nu\rho}^\lambda
      +\frac{N}{16}\omega_{\mu ab}\omega^{\mu ab}\,, & \quad TS\\[1.5mm]
      -\left(\frac18+\frac{\alpha N}{2}
      \right)R-\frac1{24}(\Gamma_{\mu\lambda}^\rho)^2+\frac{N}{24}\omega_{\mu
        ab}\omega^{\mu ab}\,, & \quad MR  \\[1.5mm]
        -\left(\frac18+\frac{\alpha N}{2} \right)R\,, & \quad DR
    \end{array}
  \right.
\end{eqnarray}
Since the process of gauging does not introduce further ambiguities than those already taken into account  in~\cite{Bastianelli:2011cc}, we conclude that the regularization there discussed is suitable for the model of the previous section, provided one sets $\alpha = -\frac18$ and
\begin{equation}\label{eq:Vtot}
V = V_{CT} -\frac{(N-2)(D+N-2)}{16(D-1)}R~.
\end{equation}
Above $TS$ refers to Time Slicing regularization~\cite{DeBoer:1995hv,deBoer:1995cb}, $MR$ refers to Mode Regularization~\cite{Bastianelli:1991be,Bastianelli:1992ct,Bastianelli:1998jm,Bastianelli:1998jb,Bonezzi:2008gs}
 and $DR$ refers to Dimensional Regularization~\cite{Kleinert:1999aq,Bastianelli:2000pt,Bastianelli:2000nm,Bastianelli:2002qw,Bastianelli:2005vk},
that are the three regularization schemes developed in the past to treat one-dimensional nonlinear sigma models (particles in curved space).
In the present work we adopt $DR$ to compute the short time perturbative expansion of~\eqref{eq:nlsm}. We parametrize the coordinates of the circle as 
$x^\mu(\tau) = x^\mu +q^\mu(\tau)  $, where $x^\mu$ is the initial/final point of the circle and $q^\mu(\tau)$ are quantum fluctuations with Dirichlet boundary conditions $q^\mu(0) = q^\mu(1) =0$. Fermions have antiperiodic boundary conditions on the circle and have no zero modes. We then expand the metric and the spin connection about the point  $x^\mu$ using Riemann normal coordinates,  and get
\begin{align}
g_{\mu\nu}(x(\tau)) &= g_{\mu\nu} + \frac13 R_{\alpha\mu\nu\beta} q^\alpha q^\beta +\frac16 \nabla_\gamma R_{\alpha\mu\nu\beta} q^\alpha q^\beta q^\gamma \nonumber\\& + R_{\alpha\beta\mu\nu\gamma\delta} q^\alpha q^\beta q^\gamma q^\delta+O(q^5)\\
\omega_{\mu ab}(x(\tau)) &= \frac12  R_{\alpha\mu ab} q^\alpha +\frac13  \nabla_\alpha R_{\beta\mu ab} q^\alpha q^\beta \nonumber\\&+\Bigl[\frac18  \nabla_\alpha \nabla_\beta R_{\gamma\mu ab} +\frac{1}{24}R^{\tau}{}_{\alpha\beta\mu} R_{\gamma\tau ab}\Bigr]q^\alpha q^\beta q^\gamma + O(q^4)
\end{align}
where $R_{\alpha\beta\mu\nu\gamma\delta} = \frac{1}{20} \nabla_\delta \nabla_\gamma R_{\alpha\mu\nu\beta} +\frac{2}{45}R_{\alpha\mu}{}^\sigma{}_\beta R_{\gamma\sigma\nu\delta}$. All the tensors are here evaluated at the initial point $x^\mu$. Above we only give the terms that are needed to obtain a perturbative expansion to order $\beta^2$.  We thus get
\begin{align}
& \Gamma[g]=K_N \int d^Dx \int_0^\infty \frac{d\beta}{\beta}\prod_{k=1}^{n}\int_0^{2\pi}\frac{d\theta_k}{2\pi} \Big({\rm Det}(\partial_\tau -\theta_{\rm vec})_{ABC} \Big)^{-1}{\rm Det'}(\partial_\tau -\theta_{\rm adj})_{PBC}\nonumber\\ &\times \int_{DBC} Dq Da Db Dc D\bar\psi D\psi D\eta ~e^{ -\frac{1}{\beta} \int_0^1d\tau
\big( \frac{1}{2} g_{\mu\nu} (\dot q^\mu \dot q^\nu + a^\mu a^\nu + b^\mu c^\nu)
  +\sum_k \bar\psi_{a k} ( \partial_\tau +i\theta_k )\psi^a_k +\frac12 \eta_a \partial_\tau \eta^a\big)} \nonumber\\[1mm]& \times e^{-S_{int}}
    \label{eq:nlsm'}
\end{align}
where we have exponentiated the reparametrization invariant measure by means of measure ghosts $a,b,c$~\cite{Bastianelli:1991be,Bastianelli:1992ct} and have complexified the $2n$ fermions $\psi$; the leftover uncomplexified Majorana fermion $\eta$ is only present when the number of supersymmetries $N$ is odd --i.e. for half-integer spin. From the quadratic part of the action one gets the path integral normalization and the propagators for all fields, that are reported in Appendix~\ref{app:propagators}, whereas higher order terms form the interacting action
\begin{align}
&S_{int} = \frac{1}{\beta} \int_0^1d\tau \nonumber\\
&\Biggl[ \Big(\frac16 R_{\alpha\mu\nu\beta} q^\alpha q^\beta +\frac{1}{12} \nabla_\gamma R_{\alpha\mu\nu\beta} q^\alpha q^\beta q^\gamma +\frac12 R_{\alpha\beta\mu\nu\gamma\delta} q^\alpha q^\beta q^\gamma q^\delta\Big)(\dot q^\mu \dot q^\nu + a^\mu a^\nu + b^\mu c^\nu)\nonumber\\
&+ \Big( \frac12  R_{\alpha\mu ab} q^\alpha +\frac13  \nabla_\alpha R_{\beta\mu ab} q^\alpha q^\beta 
\nonumber\\
&\hskip0.5cm+ \frac{1}{24}(3  \nabla_\alpha \nabla_\beta R_{\gamma\mu ab} +R^{\tau}{}_{\alpha\beta\mu} R_{\gamma\tau ab})q^\alpha q^\beta q^\gamma 
\Big)\dot q^\mu \Biggl(\sum_{k=1}^{n} \bar\psi^a_k \psi^b_k +\frac12 \eta^a \eta^b\Biggr)\nonumber\\
&+ \alpha\Big(R_{abcd} + q^\alpha \nabla_\alpha R_{abcd} + \frac12 q^\alpha q^\beta\nabla_\alpha \nabla_\beta R_{abcd} \Big)\psi^a\cdot\bar \psi^b  \Big(\psi^c\cdot\bar \psi^d +\eta^c \eta^d \Big) \nonumber\\
&+\beta^2\Big(V +q^\alpha \nabla_\alpha V +\frac12 q^\alpha q^\beta  \nabla_\alpha \nabla_\beta V \Big) \Biggl]
\end{align}
 whose path integral average is computed using the Wick theorem.  We thus get
\begin{align}
 \Gamma[g]&=\int_0^\infty \frac{d\beta}{\beta}\int \frac{d^Dx \sqrt{|g|}}{(2\pi \beta)^{D/2}} \prod_{k=1}^{n}\int_0^{2\pi}\frac{d\theta_k}{2\pi} d(\theta;D,S)~\Big< e^{-S_{int}} \Big>
    \label{eq:nlsm''}
\end{align}
with $\frac{\sqrt{|g|}}{(2\pi\beta)^{D/2}}$ being the normalization of the bosonic path integral in $D$ dimensions with Dirichlet boundary conditions, whereas the fermionic normalization contributes to the moduli integrand  
\begin{align}\label{moremeasure}
&d(\theta;D,N) = K_N \Big({\rm Det}(\partial_\tau -\theta_{\rm vec})_{ABC} \Big)^{\frac{D}{2}-1}{\rm Det'}(\partial_\tau -\theta_{\rm adj})_{PBC} \\ 
&=\left\lbrace\begin{array}{ll}
\frac{2}{2^n n!}
{\displaystyle \prod_{k=1}^n}
\Big (2 \cos\frac{\theta_k}{2} \Big )^{D-2}
{\displaystyle \prod_{k<l}} \left [
\Big (2 \cos\frac{\theta_l}{2} \Big )^2
-\Big (2 \cos\frac{\theta_k}{2} \Big )^2  \right ]^2, &\ N=2n \\[.5cm]
\frac{2^{\frac{D}{2}-1}}{2^{n} n!}
{\displaystyle \prod_{k=1}^{n}}
\Big (2 \cos\frac{\theta_k}{2} \Big )^{D-2}
\Big (2 \sin\frac{\theta_k}{2} \Big )^2  
{\displaystyle \prod_{k<l}} \left[ \Big (2 \cos\frac{\theta_l}{2} \Big )^2
-\Big (2 \cos\frac{\theta_k}{2} \Big )^2
\right]^2, &\ N=2n+1
\end{array}\right.
\nonumber
\end{align}
that integrated gives
\begin{align}
Dof(D,N) = \prod_{k=1}^{n}\int_0^{2\pi}\frac{d\theta_k}{2\pi}~d(\theta;D,N) := a_0~,
\end{align}
the number of degrees of freedom for the higher spin field described by the locally supersymmetric spinning particle model with $N$ supersymmetries~\cite{Bastianelli:2007pv}, i.e. the physical polarizations of a particle of 
spin $S=N/2$. By factoring out the number of degrees of freedom, we can finally write the above effective action in a compact way as
\begin{align} 
 \Gamma[g]&=a_0 \int_0^\infty \frac{d\beta}{\beta}\int \frac{d^Dx \sqrt{|g|}}{(2\pi \beta)^{D/2}} ~\Big<\!\! \Big< e^{-S_{int}} \Big>\!\!\Big> =\int_0^\infty \frac{d\beta}{\beta} Z(\beta)
\label{eq:nlsm3}
\end{align}
with $\big<\!\!\big<\cdots \big>\!\! \big>$ representing the average over the path integral and over the moduli space. Hence, for the effective action density in proper time we get
\begin{align}
Z(\beta) &= a_0\int \frac{d^Dx \sqrt{|g|}}{(2\pi \beta)^{D/2}}~\Big<\!\! \Big< e^{-S_{int}} \Big>\!\!\Big> = \int \frac{d^Dx \sqrt{|g|}}{(2\pi \beta)^{D/2}} \Big(a_0 +a_1\beta +a_2\beta^2+O(\beta^3) \Big)
\end{align}
and we parametrize the Seleey-DeWitt coefficients $a_i$ as follows
\begin{align}
 a_0 \Big( 1+v_2 R \beta + (v_3 R_{abcd}^2 +v_4 R_{ab}^2 +v_5 R^2 +v_6 \nabla^2 R)\beta^2+O(\beta^3) \Big)~. 
\end{align}
Next we compute the numerical coefficients $v_i$.

\section{Heat kernel expansion for higher spin fields in (A)dS}
\label{sec:sdw-ads}
Equipped with the results of the previous sections we can now compute
the heat kernel in a perturbative expansion for higher spin fields on (A)dS spaces, using
the $O(N)$ spinning particle representation discussed above. Although
in the previous sections we gauge fixed the locally
supersymmetric action for maximally symmetric spaces only, here
we compute the expansion keeping an unspecified metric in the sigma model and only at the end of
the section will we specialize to (A)dS spaces.
This we do mostly for future convenience, as intermediate results might be useful when considering more general spacetimes,
such as the conformally flat spaces.
Since in the following we adopt dimensional regularization, the total potential acquires the form:
$$
V=w\,R\,,\,\,\,\,\,\,\,\,\,\textrm{with}\,\,\,\,w(D,N,\alpha):=w_{CT}(N,\alpha)+w_{(A)dS}(D,N)
$$
where
\begin{align}
w_{CT}(N,\alpha) &= -\Big(\frac18 +\frac{\alpha N}{2} \Big)\,,\quad
w_{(A)dS}(D,N) = -\frac{(N-2)(D+N-2)}{16(D-1)}~,
\end{align}
as follows from~(\ref{eq:VCT},\ref{eq:Vtot}).

\subsection{Integer spins} 
For this case we set $N=2n$.  One can complexify fermions and, with
the help of propagators given in Appendix~\ref{app:propagators}, one gets for the perturbative average
\begin{align}
    \Big \langle e^{-S_{int}} \Big \rangle &= \exp\Biggl\{ -\beta\Biggl[
    \frac1{24} +\alpha \left(n-\sum_k \cos^{-2}\frac{\theta_{k}}{2}
      \right)+w\Biggr]\,R\nonumber\\
    &+\beta^2 \Biggl[ -\frac1{720}
    R_{\alpha\beta}^2+\left(\frac1{720}
      -\frac1{192}\sum_k \cos^{-2}\frac{\theta_{k}}{2}\right)R_{\alpha\mu\nu\beta}^2
    -\left(\frac1{480}+\frac w{12}\right) \nabla^2 R\Biggr]\nonumber \\
    & -\frac{\alpha\beta^2}{12} \left(n-\sum_k \cos^{-2}\frac{\theta_{k}}{2}
    \right) \nabla^2 R\nonumber \\
    & +(\alpha \beta)^2 \Biggl[\Biggl(
    \left(\sum_k \cos^{-2}\frac{\theta_{k}}{2}\right)^2-\frac12 \sum_k
    \cos^{-4}\frac{\theta_{k}}{2}\Biggr)R_{\alpha\mu\nu\beta}^2\nonumber \\ &\qquad\qquad +2\Biggl(
    \sum_k \cos^{-2}\frac{\theta_{k}}{2}-\sum_k
    \cos^{-4}\frac{\theta_{k}}{2}\Biggr) R_{\alpha\beta}^2\Biggr]+O(\beta^3)\Biggr\}~,
\end{align}
that, for $\alpha=-1/8$ reduces to
\begin{align}
   \Big< e^{-S_{int}} \Big>
     &=1-\beta \left(\frac{1-3n}{24}+\frac{1}{8}\sum_{k}\cos^{-2}\frac{\theta_{k}}{2}
       +w\right)R\nonumber \\
     &+\beta^{2}\Biggl\{
     \frac12
     \left(\frac{1-3n}{24}+\frac{1}{8}\sum_{k}\cos^{-2}\frac{\theta_{k}}{2}+w\right)^2\!\!\!R^2
     \nonumber\\
     &\qquad
     +\left(-\frac{1}{720}-\frac{1}{32}\sum_{k}\cos^{-4}\frac{\theta_{k}}{2}+\frac{1}{32}\sum_{k}
       \cos^{-2}\frac{\theta_{k}}{2}\right)R^{2}_{ab}\displaybreak
       \nonumber\\
     &\qquad
     +\left(\frac{1}{720}-\frac{1}{192}\sum_{k}\cos^{-2}\frac{\theta_{k}}{2}
       +\frac{1}{64}\left(\sum_{k}\cos^{-2}\frac{\theta_{k}}{2}\right)^2\right.\nonumber\\
       &\left.\qquad-\frac{1}{128}\sum_{k}\cos^{-4}\frac{\theta_{k}}{2}\right)R^{2}_{abcd}
     \left.-\left(\frac{1-5n}{480}+\frac{1}{96}\sum_{k}\cos^{-2}\frac{\theta_{k}}{2}
     +\frac{w}{12}\right)\nabla^{2}R\right\}\nonumber \\&\qquad+O(\beta^3)\,,
\label{MasterNeven}
\end{align}
with
$$w=w(D=2d,N=2n,\alpha=\scalebox{.95}{$-\frac 1 8$})=-\frac{(N-2)(N-1)}{16(D-1)}=-\frac{(2n-1)(n-1)}{8(2d-1)}\,.$$
We are ready now to extract the Seeley-DeWitt coefficients for arbitrary integer spin  $S=n$ in arbitrary even dimension $D=2d$; 
to this aim we integrate \eqref{MasterNeven} against the modular measure given in (\ref{eq:nlsm''},\ref{moremeasure}) and get:
\begin{equation}\label{sdweven}
a_0=\left\{\begin{array}{ll}
1\,, & n=0\\[2mm]
{\displaystyle 2^{n-1}\frac{(2d-2)!}{[(d-1)!]^2}\prod_{k=1}^{n-1}\frac{k(2k-1)!(2k+2d-3)!}{(2k+d-2)!(2k+d-1)!} }\,,\ & n> 0
\end{array}\right.
\end{equation}
and
\begin{align}
v_2&=\frac{3n-1}{24}-\frac 1 8 \,\I_1-w\nonumber\\[2mm]
v_3&=\frac{1}{720}-\frac{n(n+1)}{256}+\frac{3n+1}{384}\,\I_1+\frac 3{256}\,\I_2+\frac 1{256}\,\I_3\nonumber\\[2mm]
v_4&=-\frac 1 {720}+\frac{n(n+1)}{64}-\frac n{32}\,\I_1+\frac 1{64}\,\I_2-\frac 1{64}\,\I_3\nonumber\\[2mm]
v_5&=\frac 1 2 \left(\frac{9n^2-21n+2}{1152}-\frac{w(3n-1)}{12}+w^2\right)+\frac{1}{2}\left(\frac{5-3n}{192}+\frac{w}{4}\right)\I_1+\frac{1}{256}\big(\I_2+\I_3\big)\nonumber\\[2mm]
v_6&=\frac{5n-1}{480}-\frac w {12}-\frac{1}{96}\,\I_1
\label{sdweven1}
\end{align}
with
\begin{equation}
\begin{split}
\I_1&=\frac{2n(n+d-2)}{2d-3}\\[2mm]
\I_2&=\frac{4n(n-1)(n+d-1)(n+d-2)}{(2d-3)(2d-1)}\\[2mm]
\I_3&=\frac{n(n+1)(4n^2-1)}{(2d-3)(2d-5)}
\end{split}
\end{equation}
Detailed computation of modular integrals is given in Appendix~\ref{KANEKO}.  Let us now briefly comment on the results described above  in~(\ref{sdweven},\ref{sdweven1}):\\
$\bullet$ For $n=0$, the formalism describes a conformally coupled scalar field and the expected results are easily obtained.\\
$\bullet$ For $n=1$, (\ref{sdweven},\ref{sdweven1}) reproduce the well known Seeley-DeWitt coefficients for a degree $(d-1)$ differential form (vector field in $D=4$) \cite{Bastianelli:2005vk} on a general background.\\
$\bullet$ For $n\ge 2$, the spinning particle consistently propagates on conformally flat manifolds. However, for this case, in the previous sections we limited the computation of  the BRST charge to (A)dS spaces. Hence the structure of the Seeley-DeWitt coefficients  reduces to
$$
a_0\,\Big(1+v_2R\beta+vR^2 \beta^2\Big)\qquad\textrm{with}\qquad v=\frac{1}{d(2d-1)}v_3+\frac 1{2d} v_4+v_5\,.
$$
\paragraph{\underline{Example:} $D=4\,,\ {\rm spin}\ n$\\}
In 4-dimensional space-time the model describes completely symmetric tensors of spin $n$, and the Seeley-DeWitt coefficients are given by:
\begin{equation}
\begin{split}
a_0&=\left\{\begin{array}{ll}
1\,,\ & n=0\\
2\,, & n>0
\end{array}\right.
\,,\,\,\quad v_2=-\frac{n^2}{6}\,,\,\,\quad v_3=\frac{1}{720}-\frac{n^2}{96}\\[2mm]
v_4&=-\frac 1 {720}-\frac{n^2}{48}+\frac{n^4}{12}\,,\,\,v_5=\frac{1}{96}n^2-\frac 1{36}n^4\,,\,\,
v_6=\frac{1}{720}-\frac 1{72}n^2\;,
\end{split}
\end{equation}
When $n\ge2$ the restriction to (A)dS yields:
\begin{equation}
v=\frac 1 6 v_3+\frac 1 4 v_4+v_5=-\frac{1}{8640}+\frac{1}{288}\,n^2-\frac{1}{144}\,n^4
\end{equation}
We again recognize for $n=0,1$ the known coefficients for a conformally improved scalar  and an ordinary spin one vector field. For $n >0$ the first coefficient $a_0$ represents the two polarizations of massless particles of spin $n$.

The case of $n=2$ corresponds to a linearized graviton on a fixed background, 
but this is true only in $D=4$. In other dimensions one has a different field content compatible with conformal invariance.

\subsection{Half-integer spins} In such a case one can only complexify $2n$
fermions. The left-over one has no $\theta$, and one thus gets
\begin{align}
    \Big \langle e^{-S_{int}} \Big \rangle &= \exp\Biggl\{ -\beta\Biggl[
    \frac1{24}+w +\alpha \left(n-\sum_k \cos^{-2}\frac{\theta_{k}}{2}
      \right)\Biggr]R 
      \nonumber\\
    &+\beta^2 \Biggl[ -\frac1{720}
    R_{\alpha\beta}^2-\left(\frac{7}{5760}
      +\frac1{192}\sum_k \cos^{-2}\frac{\theta_{k}}{2}\right)R_{\alpha\mu\nu\beta}^2
    -\left(\frac1{480}+\frac w {12}\right) \nabla^2 R\Biggr] \displaybreak\nonumber\\
    & -\frac{\alpha\beta^2}{12} \left(n-\sum_k \cos^{-2}\frac{\theta_{k}}{2}
    \right) \nabla^2 R\nonumber \\
    & +(\alpha \beta)^2 \Biggl[\Biggl(
    \left(\sum_k \cos^{-2}\frac{\theta_{k}}{2}\right)^2+\sum_k
    \cos^{-2}\frac{\theta_{k}}{2}-\frac12 \sum_k
    \cos^{-4}\frac{\theta_{k}}{2}\Biggr)R_{\alpha\mu\nu\beta}^2\nonumber\\ &\qquad\qquad +2\Biggl(
    \sum_k \cos^{-2}\frac{\theta_{k}}{2}-\sum_k
    \cos^{-4}\frac{\theta_{k}}{2}\Biggr)R_{\alpha\beta}^2 \Biggr]+O(\beta^3)\Biggr\}~,
\end{align}
that, for $\alpha=-1/8$, reduces to
\begin{align}
     \Big< e^{-S_{int}} \Big>
     &=1-\beta\left( \frac{1-3n}{24}+\frac{1}{8}\sum_{k}\cos^{-2}\frac{\theta_{k}}{2}+w
       \right)R\nonumber\\
     &+\beta^{2}\Biggl\{
     \frac12
     \left(\frac{1-3n}{24}+\frac{1}{8}\sum_{k}\cos^{-2}\frac{\theta_{k}}{2}+w\right)
       ^2R^2
       \nonumber\\
     &\qquad
     +\left(-\frac{1}{720}-\frac{1}{32}\sum_{k}\cos^{-4}\frac{\theta_{k}}{2}+\frac{1}{32}\sum_{k}\cos^{-2}
       \frac{\theta_{k}}{2}\right)R^{2}_{ab}\nonumber\\
     &\qquad
     +\left(-\frac{7}{5760}+\frac{1}{96}\sum_{k}\cos^{-2}\frac{\theta_{k}}{2}
       +\frac{1}{64}\left(\sum_{k}\cos^{-2}\frac{\theta_{k}}{2}\right)^2\right.
    \nonumber \\
     &\qquad\left.-\frac{1}{128}\sum_{k}\cos^{-4}\frac{\theta_{k}}{2}\right)R^{2}_{abcd}
     -
     \left(\frac{1-5n}{480}+\frac{w}{12}+\frac{1}{96}\sum_{k}\cos^{-2}\frac{\theta_{k}}{2}
     \right)\nabla^{2}R \Biggr\}\nonumber\\ &\qquad
    +O(\beta^3)
   \end{align}
where now we use
$$w=w(D=2d,N=2n+1,\alpha=\scalebox{.95}{$-\frac 1 8$})=-\frac{(N-2)(N-1)}{16(D-1)}=-\frac{n(2n-1)}{8(2d-1)}\,.$$
We compute, in analogy with the previous section, the Seeley-DeWitt coefficients for arbitrary half-integer spin $S=n+\frac 1 2$ in arbitrary even dimension $2d$, represented by spinor-tensors corresponding to potentials with rectangular Young tableaux of $n$ columns and $d-1$ rows; we get:
\begin{equation}\label{sdwodd}
\begin{split}
a_0&=\frac{2^{d-2+n}}{d}\frac{(2d-2)!}{[(d-1)!]^2}\prod_{k=1}^{n-1}\frac{(k+d-1)(2k+1)!(2k+2d-3)!}{(2k+d-1)!(2k+d)!} 
\end{split}
\end{equation}
and
\begin{align}
v_2&=\frac{3n-1}{24}-\frac 1 8 \,\It_1-w \displaybreak
\nonumber\\[2mm]
v_3&=-\frac{7}{5760}-\frac{n(n+1)}{256}+\frac{3n+7}{384}\,\It_1+\frac 3{256}\,\It_2+\frac 1{256}\,\It_3\nonumber\\[2mm]
v_4&=-\frac 1 {720}+\frac{n(n+1)}{64}-\frac{n}{32}\,\It_1+\frac 1{64}\,\It_2-\frac 1{64}\,\It_3\nonumber\\[2mm]
v_5&=\frac 1 2 \left(\frac{9n^2-21n+2}{1152}-\frac{w(3n-1)}{12}+w^2\right)+\frac{1}{2}\left(\frac{5-3n}{192}+\frac{w}{4}\right)\It_1+\frac{1}{256}\big(\It_2+\It_3\big)\nonumber\\[2mm]
v_6&=\frac{5n-1}{480}-\frac w {12}-\frac{1}{96}\,\It_1\label{sdwodd1}
\end{align}
with
\begin{equation}
\begin{split}
\It_1&=\frac{2n(n+d-1)}{2d-3}\\[2mm]
\It_2&=\frac{4n(n-1)(n+d-1)(n+d)}{(2d-3)(2d-1)}\\[2mm]
\It_3&=\frac{n(n+1)(2n+1)(2n+3)}{(2d-3)(2d-5)}\,.
\end{split}
\end{equation}
The modular integrals are again computed in details in Appendix~\ref{KANEKO}.

In the half-integer spin case the spinning particle model we start with is consistent on any background only if $n=0$ (\emph{i.e.} spin $\frac 1 2$).
When $n\ge 1$ we restrict our analysis to (A)dS spaces and at order $\beta^2$ in the expansion of the effective action the only term that survives is $a_0vR^2$ where, again, $v=\frac{1}{d(2d-1)}v_3+\frac 1{2d} v_4+v_5$. 
\paragraph{\underline{Example:} $D=4\,,\ {\rm spin}\ n+\frac12$\\}
In 4-dimensional space-time we describe spinor-tensors with $n$ completely symmetric vector indices and one spinor index \big(\emph{i.e.} spin $n+\frac 1 2$\big). The Seeley-DeWitt coefficients we find are: 
\begin{align}
a_0&=2\,,\,\,\quad v_2=-\frac{(2n+1)^2}{24}\,,\,\,\quad v_3=-\frac{7}{5760}-\frac{n}{96}-\frac{n^2}{96} \nonumber\\[2mm]
v_4&=-\frac 1 {720}+\frac{n}{48}+\frac{5n^2}{48}+\frac{n^3}{6}+\frac{n^4}{12}\,,\,\,v_5=\frac{1}{1152}+\frac{1}{144}n-\frac{5}{96}n^2-\frac{19}{288}n^3-\frac 1{144}n^4\,,\,\,
\nonumber\\[2mm]
v_6&=-\frac{1}{480}-\frac 1{72}n-\frac 1{72}n^2 \,.
\end{align}
When $n=0$ the previous formulas reproduce the well know Seeley-DeWitt coefficients for a spinor field~\cite{Bastianelli:2002qw}, while for $n\ge 1$ in (A)dS we get:
$$
v=\frac{11}{34560}+\frac{n}{96}-\frac{n^2}{36}-\frac{7n^3}{288}+\frac{n^4}{72}\,.
$$
Let us stress again that in 4 dimension we recognize in $a_0=2$ the two polarizations of a massless half-integer spin field.
\vfill\eject

\acknowledgments{The work of FB was supported in part by the MIUR-PRIN contract 2009-KHZKRX.  The work of OC  was partly funded by SEP-PROMEP/103.5/11/6653. 
EL acknowledges partial support of SNF Grant No. 200020-131813/1.
OC and EL are grateful to the Dipartimento di Fisica and INFN Bologna for hospitality and support while parts of this work were completed.}

\appendix

\section{Hamiltonian BRST quantization}
\label{app:brst}
The hamiltonian BRST formalism is a construction that allows to
convert the local (gauge) symmetry of the unfixed action (in
hamiltonian form) to a global symmetry of the gauge-fixed action. It
makes use of the double aspect that first-class generators have, as
restrictions on the phase-space and generators of gauge
transformations (see for examples~\cite{Henneaux:1992ig}).

One defines a differential $\delta$ (the Koszul-Tate differential)
that acts as a derivative in the directions orthogonal to the
constrained phase-space manifold and is nilpotent, $\delta^2=0$.
Hence the definition
\begin{eqnarray}
  \delta z^\alpha =0,\quad z^\alpha=(p_\mu,x^\mu,\psi_i^a)~.
\end{eqnarray}
Moreover, one extends the phase space defining ghosts ${\cal C}^A$ and ghost momenta
${\cal P}_A$, such that $\{{\cal P}_A, {\cal C}^B\} = -\delta_A^B$ and
\begin{eqnarray}
  \delta {\cal C}^A =0,\quad \delta {\cal P}_A =-G_A
\end{eqnarray}
with $G_A$ first class constraints. The operator $\delta$ thus defines a natural grading, characterized by
the antighost number
\begin{eqnarray}
  \overline{\rm gh} (\delta) =-1,\quad \overline{\rm gh} (z) =0=\overline{\rm
    gh} ({\cal C}),\quad \overline{\rm gh} ({\cal P})=1~.
\end{eqnarray}
Note that the bracket itself in the ghost sector has antighost number $-1$.
Another grading is the Grassmann parity
\begin{eqnarray}
  \varepsilon_A := \varepsilon(G_A)
\end{eqnarray}
so that, since $\varepsilon(\delta)=1$, we have
\begin{eqnarray}
  \varepsilon({\cal C}^A) =\varepsilon({\cal P}_A)
  =\varepsilon_A+1,\quad {\rm mod\ }2~.
\end{eqnarray}
One also introduces another derivative $d$ that acts parallel to the gauge
orbits. It is defined on functions of the original phase space, $\phi(z)$, as
\begin{eqnarray}
  d\phi = \{\phi,{\cal C}^A G_A\} = \{\phi, G_A\} {\cal C}^A,\quad
  \overline{\rm gh} (d) =0,\ \varepsilon(d) =1~.
\end{eqnarray}
Finally one seeks a differential $s$ that is a graded sum of $\delta$,
$d$ and higher order (in antighost number) derivatives, such that it results
nilpotent on the extended phase space involving ghosts
\begin{eqnarray}
  s=\delta + d + `` {\rm higher\ order\ terms} `` \,,\quad s^2 =0~.
\end{eqnarray}
Thanks to antighost grading, nilpotency of $s$ implies
\begin{eqnarray}
  && \delta^2 = 0\\
  && d\delta + \delta d =0\label{d-delta}\\
  && d^2 = -\{\delta,\Delta\}\label{d2}\\
  && {}\cdots{}\nonumber
\end{eqnarray}
Equations~\eqref{d-delta},\eqref{d2} mean that $d$ is a ``differential modulo
$\delta$''. The first one is satisfied, along with the grading
properties, if one defines the following rules for the action of $d$
on the extended phase space
\begin{eqnarray}
  d{\cal P}_A = (-)^{\varepsilon_A} {\cal C}^C F_{C
  A}^B {\cal P}_B\,,\quad d{\cal C}^A=0
\end{eqnarray}
where $F$'s are structure functions and only depend upon the original
phase space variables
\begin{eqnarray}
  \big\{G_A,G_B\big\} = F_{AB}^C
  G_{C}\,,\quad  F_{AB}^C=F_{AB}^C(z)~.
\end{eqnarray}
One then seeks a BRST operator $\Omega$
\begin{eqnarray}
  \Omega =\sum_{p\geq 0} \stackrel{_{(p)}}{\Omega}\,,\quad \overline{\rm
  gh}\big(\stackrel{_{(p)}}{\Omega}\big) =p
\end{eqnarray}
that implements the action of the differential $s$ as
\begin{eqnarray}
  s \Phi =\{\Phi,\Omega\}
\end{eqnarray}
with $\Phi(z,{\mathcal C},{\mathcal P})$ a function of the extended phase space variables, where
\begin{eqnarray}
  \stackrel{_{(0)}}{\Omega} = {\mathcal C}^A G_A
\end{eqnarray}
so that
\begin{eqnarray}
  \delta \Phi =\Big\{\Phi,\stackrel{_{(0)}}{\Omega}\Big\}_{_{{\cal C}\, {\cal P}}}
  =\{\Phi,{\cal C}^A\} G_A
  \label{Q0}
\end{eqnarray}
with the lowerscript ${{\cal C} {\cal P}}$ meaning that the bracket is only
taken in the ghost sector. It is trivial to check that~\eqref{Q0}
works correctly on the extended phase space variables. For a function of the original phase space we obviously have
$ d \phi(z) =\Big\{\phi,\stackrel{_{(0)}}{\Omega}\Big\}_{_{\rm orig}}
  =\{\phi,G_A\} {\cal C}^A$. Finally, thanks to the Jacobi identity, the nilpotency conditions turns into
\begin{eqnarray}
 \Big \{\Omega, \Omega \Big\} =0~.
  \label{Q2}
\end{eqnarray}
Higher order operators have the form
\begin{eqnarray}
  \stackrel{_{(p)}}{\Omega} = {\cal C}^{B_1}\cdots {\cal C}^{B_{p+1}}
  U_{B_1\cdots B_{p+1}}^{A_1\cdots A_p} {\cal
    P}_{A_1}\cdots {\cal P}_{A_p}\,,\quad U=U(z)
\end{eqnarray}
so that the nilpotency equation~\eqref{Q2}, with the help
of~\eqref{Q0}, allows to write
\begin{eqnarray}
  && \delta \stackrel{_{(0)}}{\Omega} =0\\
  && \delta \stackrel{_{(p+1)}}{\Omega} +\frac12\Biggl( \sum_{k=0}^p
  \Big\{\stackrel{_{(p-k)}}{\Omega},\stackrel{_{(k)}}{\Omega}\Big\}_{\rm orig} +
  \sum_{k=0}^{p-1}
  \Big\{\stackrel{_{(p-k)}}{\Omega},\stackrel{_{(k+1)}}{\Omega}\Big\}_{_{{\cal C}\, {\cal
        P}}} \Biggr)=0\,,\quad p\geq 0~.
\end{eqnarray}
For example, it is easy to find the next-to-leading operator $\stackrel{_{(1)}}{\Omega}$ as
\begin{eqnarray}
  \delta \stackrel{_{(1)}}{\Omega} = -\frac12
  \Big\{\stackrel{_{(0)}}{\Omega},\stackrel{_{(0)}}{\Omega}\Big\}_{\rm
    orig}=\frac12(-)^{\varepsilon_A} {\cal C}^A {\cal C}^B F_{BA}^C G_C
  = \delta\left(- \frac12(-)^{\varepsilon_A}
  {\cal C}^A {\cal C}^B F_{BA}^C {\cal P}_C\right)\quad
\end{eqnarray}
so that, modulo a $\delta$-exact term
\begin{eqnarray}
  \stackrel{_{(1)}}{\Omega} = - \frac12(-)^{\varepsilon_A}
  {\cal C}^A {\cal C}^B F_{BA}^C {\cal P}_C~.
\end{eqnarray}
One thus recursively fixes all other terms in the graded expansion. If the constraint algebra is linear (i.e. it is a Lie algebra)
the expansion stops at $p=1$.

The gauge fixed action in hamiltonian form reads
 \begin{eqnarray}
S_{gf} = \int_0^1 dt \biggl[ \frac12 \omega_{\alpha \beta}\dot z^\alpha z^\beta + \dot {\cal C}^A {\cal P}_A -H_{BRST} -\Big\{K,\Omega\Big\}\biggr]
\end{eqnarray}
where $\omega_{\alpha \beta}$ is the symplectic matrix in canonical coordinates,
$H_{BRST}$ is the BRST-invariant extension of the extended hamiltonian, and $K$ an arbitrary gauge-fixing
fermion. This action is BRST invariant for any $K$. An important example concerns algebraic gauges for which
the gauge fields are fixed to $\hat E^A$: in such a special case
\begin{eqnarray}
  K = -\hat E^A {\cal P}_A
\end{eqnarray}
for which
\begin{eqnarray}
  \Big\{K,\Omega\Big\} = \hat E^A
  G_A-(-)^{\varepsilon_A}\hat E^A {\cal C}^B
  F_{BA}^C {\cal P}_C+\cdots\cdots~.
\end{eqnarray}
The above technique to construct the BRST charge $\Omega$ is known as Koszul-Tate resolution. 

Below we use the Koszul-Tate resolution to study an interesting class of non lie rank 3 superalgebra, and to construct the gauge fixed action for $O(N)$ spinning particles propagating on (A)dS target spaces. We do it directly at the quantum level where Poisson brackets are replaced by (anti-)commutators, such as $[{\mathcal P}_A, {\mathcal C}^B\}=-i\delta_A^B$ and ${\mathcal P}_A$ are taken to be (anti-)hermitian when (anti-)commuting whereas ${\mathcal C}^A$ are always hermitians. The master formula~\eqref{Q2} is now a nilpotency condition on the BRST charge, $\Omega^2=0$. We thus have
\begin{align}
\stackrel{_{(0)}}{\Omega} = {\cal C}^A G_A\,,\quad  \stackrel{_{(1)}}{\Omega} =\frac{i}{2}(-)^{\varepsilon_A} {\cal C}^A {\cal C}^B F_{BA}^C {\cal P}_C\,,\dots~.
\end{align}
  The hamiltonian operator is given by  ${\mathbb H}_{qu} = H_{BRST} - i\{ K,\Omega\}$, with $H_{BRST}$ a BRST-invariant hamiltonian and $K$ a gauge-fixing fermion. 

\section{Propagators}
\label{app:propagators}
Propagators are obtained by inverting the differential operators appearing in the quadratic action $ \frac{1}{\beta} \int_0^1d\tau
\big( \frac{1}{2} g_{\mu\nu} (\dot q^\mu \dot q^\nu + a^\mu a^\nu + b^\mu c^\nu)
  +\sum_k \bar\psi_{a k} ( \partial_\tau +i\theta_k )\psi^a_k +\frac12 \eta_a \partial_\tau \eta^a\big)$
\begin{eqnarray}
  \label{eq:x-prop}
  \big< q^\mu (\tau) q^\sigma (\sigma) \big > &=& -\beta g^{\mu\nu}
   \Delta(\tau,\sigma)\\
  \label{eq:a-prop}
  \big< a^\mu (\tau) a^\sigma (\sigma) \big > &=& \beta g^{\mu\nu}
   \Delta_{gh}(\tau,\sigma)\\
  \label{eq:b-prop}
  \big< b^\mu (\tau) c^\sigma (\sigma) \big > &=& -2\beta g^{\mu\nu}
   \Delta_{gh}(\tau,\sigma)\\
  \label{eq:psi-prop}
  \big< \psi^a_k (\tau) \bar \psi^b_{k'} (\sigma) \big > &=& \beta
  \delta_{kk'} \delta^{ab}
  \Delta_{AF}(\tau-\sigma,\theta_k)\\
  \label{eq:eta-prop}
  \big< \eta^a (\tau) \eta^b (\sigma) \big > &=& \beta
  \delta^{ab} \Delta_{AF}(\tau-\sigma,0)
\end{eqnarray}
with
\begin{eqnarray}
  \label{eq:delta}
  \Delta(\tau,\sigma) &=& (\tau-1)\sigma
  \theta(\tau-\sigma)+(\sigma-1)\tau \theta(\sigma-\tau)\\
\Delta_{gh}(\tau,\sigma) &=& {}^{\bullet\bullet} \Delta(\tau,\sigma) =\delta(\tau,\sigma)
\end{eqnarray}
and
\begin{eqnarray}
  \label{eq:deltaAF}
  \Big(\partial_\tau+i\theta_k\Big)\Delta_{AF}(\tau-\sigma,\theta_k) = \delta_A(\tau-\sigma)
\end{eqnarray}
that yields
\begin{eqnarray}
  \label{eq:deltaAF'}
  \Delta_{AF}(\tau-\sigma,\theta_k) = \frac{e^{-i\theta_k
      (\tau-\sigma)}}{2\cos \frac{\theta_k}{2}}
  \biggl(e^{i\theta_k/2}\theta(\tau-\sigma)-e^{-i\theta_k/2}\theta(\sigma-\tau)
  \biggr)~.
\end{eqnarray}
Hence
\begin{eqnarray}
  \label{eq:deltaAF0}
  \Delta_{AF}(0,\theta_k) &=& \frac{i}{2}\tan\frac{\theta_k}{2}\\
  \label{eq:deltaAF0'}
  \Delta_{AF}(\tau-\sigma,0) &=& \frac{1}{2}\epsilon(\tau-\sigma)~.
\end{eqnarray}

\section{Modular integrals}\label{KANEKO}

In this appendix we are going to show the detailed calculation of the modular integrals required to find the Seleey-DeWitt (SDW) coefficients presented in section \ref{sec:sdw-ads}.
We will always consider even dimensional spacetime with $D=2d$, and we shall distinguish the two cases of even and odd $N$, although the techniques will be the same.

\subsection{Even {\it N}}

We compute the modular integrals for the even $N=2n$ case.
First of all, we define the modular average of an arbitrary function $f(\theta_j)$ of the moduli $\theta_j$; by using the measure given in \eqref{eq:nlsm''} and \eqref{moremeasure}, and taking into account that modular integrals are even under $\theta_i \to 2\pi -\theta_i$,  we have:
\begin{equation}\label{modular average defined}
\left\langle\!\left\langle f(\theta_j)\right\rangle\!\right\rangle_E:=\frac1{a_0}\frac{2}{n!}\prod_{i=1}^n\int_0^\pi\frac{d\theta_i}{2\pi}\left(2\cos\frac{\theta_i}{2}\right)^{D-2}
\prod_{k<l}\left[\left(2\cos\frac{\theta_k}{2}\right)^2-\left(2\cos\frac{\theta_l}{2}\right)^2\right]^2f(\theta_j)
\end{equation}
where $a_0$ is the normalization factor giving the degrees of freedom, that ensures $\left\langle\!\left\langle1\right\rangle\!\right\rangle_E=1$, and reads
\begin{equation}\label{a_0}
a_0:=\frac{2}{n!}\prod_{i=1}^n\int_0^\pi\frac{d\theta_i}{2\pi}\left(2\cos\frac{\theta_i}{2}\right)^{D-2}
\prod_{k<l}\left[\left(2\cos\frac{\theta_k}{2}\right)^2-\left(2\cos\frac{\theta_l}{2}\right)^2\right]^2\;.
\end{equation}
The result  for \eqref{a_0} is already known from \cite{Bastianelli:2007pv}, but will be rederived  here.
Since all the integrals we need will be expressed as generalizations of the Selberg's integral, it is convenient to change variables as $x_i=\sin^2\frac{\theta_i}{2}$, ranging from $0$ to $1$. The average of a function $f(x_j):= f(\theta(x_j))$ becomes
\begin{equation}\label{modular average x var}
\left\langle\!\left\langle f(x_j)\right\rangle\!\right\rangle_E:=\frac{\mathcal{N}}{a_0}\prod_{i=1}^n\int_0^1dx_i\,x_i^{-1/2}(1-x_i)^{d-3/2}
\prod_{k<l}(x_k-x_l)^2f(x_j)\;,
\end{equation}
where
\begin{equation}\label{Norm even N}
\mathcal{N}=\frac{2^{2(d-1)n+(n-1)(2n-1)}}{\pi^nn!}\;.
\end{equation}

The averages we need to compute can be read down from \eqref{MasterNeven}, and are
\begin{equation}\label{angular averages}
\begin{split}
\textbf{I}_1 &:= \left\langle\!\!\!\left\langle\sum_{i=1}^n\cos^{-2}\frac{\theta_i}{2}\right\rangle\!\!\!\right\rangle_E=
\left\langle\!\!\!\left\langle\sum_{i=1}^n\frac{1}{1-x_i}\right\rangle\!\!\!\right\rangle_E\;,\\[2mm]
\textbf{J} &:= \left\langle\!\!\!\left\langle\sum_{i,j=1}^n\cos^{-2}\frac{\theta_i}{2}\cos^{-2}\frac{\theta_j}{2}\right\rangle\!\!\!\right\rangle_E=
\left\langle\!\!\!\left\langle\sum_{i,j=1}^n\frac{1}{(1-x_i)(1-x_j)}\right\rangle\!\!\!\right\rangle_E\;,\\[2mm]
\textbf{K} &:= \left\langle\!\!\!\left\langle\sum_{i=1}^n\cos^{-4}\frac{\theta_i}{2}\right\rangle\!\!\!\right\rangle_E=
\left\langle\!\!\!\left\langle\sum_{i=1}^n\frac{1}{(1-x_i)^2}\right\rangle\!\!\!\right\rangle_E\;.
\end{split}
\end{equation}
For notational convenience we gave the names $\textbf{J}$ and $\textbf{K}$ to the corresponding averages, since they will be found as linear combinations of other quantities named $\textbf{I}_2$ and $\textbf{I}_3$, in terms of which the SDW coefficients are presented in the paper.

Let us focus now on the factor $a_0$, that gives the degrees of freedom of the model. In the $x_i$ variables it is given by
\begin{equation}\label{a0 x var}
a_0=\mathcal{N}\prod_{i=1}^n\int_0^1dx_i\,x_i^{-1/2}(1-x_i)^{d-3/2}\prod_{k<l}(x_k-x_l)^2\;.
\end{equation}
There is a well known result by Selberg \cite{metha,kaneko} for such kind of integrals, that gives:
\begin{equation}\label{Selberg}
S_n(\alpha,\beta):=\prod_{i=1}^n\int_0^1dx_i\,x_i^{\alpha}(1-x_i)^{\beta}\prod_{k<l}(x_k-x_l)^2=
\prod_{k=1}^n\frac{k!\Gamma(k+\alpha)\Gamma(k+\beta)}{\Gamma(k+n+\alpha+\beta)}\;,
\end{equation}
from which we obtain, after inserting the factor \eqref{Norm even N} and rearranging the product in \eqref{Selberg}:
\begin{equation}\label{a0 computed}
a_0=\mathcal{N}\,S_n(-\tfrac12,d-\tfrac32)=2^{n-1}\frac{(2d-2)!}{[(d-1)!]^2}\prod_{k=1}^{n-1}\frac{k(2k-1)!\,(2k+2d-3)!}{(2k+d-2)!\,(2k+d-1)!}\;,
\end{equation}
that indeed coincides with the result found in \cite{Bastianelli:2007pv}.

To proceed further, let us consider the following generalization of Selberg's integral by Aomoto \cite{kaneko,aomoto}:
\begin{equation}\label{Sn1}
\begin{split}
S_{n,1}(\alpha,\beta;t)&:= \prod_{i=1}^n\int_0^1dx_i\,x_i^{\alpha}(1-x_i)^{\beta}(x_i-t)\prod_{k<l}(x_k-x_l)^2\\
&=S_n(\alpha,\beta)\,\frac{n!}{{\displaystyle \prod_{k}}(k+n+\alpha+\beta)}\,P_n^{(\alpha,\beta)}(1-2t)\;,
\end{split}
\end{equation}
where $P_n^{(\alpha,\beta)}(1-2t)$ is the Jacobi polynomial of degree $n$. By taking a derivative of \eqref{Sn1} with respect to $t$, and evaluating it at $t=1$ we get very close to the definition of $\textbf{I}_1$, and precisely we have
\begin{equation}\label{I1}
\textbf{I}_1=\frac{\mathcal{N}}{a_0}\,(-)^n\,\de_tS_{n,1}(-\tfrac12, d-\tfrac52;t)\rvert_{t=1}=(-)^n\,\frac{\de_tS_{n,1}(-\tfrac12, d-\tfrac52;t)\rvert_{t=1}}{S_n(-\tfrac12,d-\tfrac32)}\;.
\end{equation}
The basic properties of Jacobi polynomials that we need for such calculation are:
\begin{equation}\label{jacobi pol}
\begin{split}
\frac{d^k}{dz^k}P_n^{(\alpha,\beta)}(z) &= \frac{\Gamma(\alpha+\beta+n+1+k)}{2^k\Gamma(\alpha+\beta+n+1)}\,P^{(\alpha+k,\beta+k)}_{n-k}(z)\;,\\
P_n^{(\alpha,\beta)}(-1)&=(-)^n\binom{n+\beta}{n}\;.
\end{split}
\end{equation}
We can now compute $\textbf{I}_1$ by inserting \eqref{Sn1} into \eqref{I1}, and using the relations \eqref{jacobi pol} and the result \eqref{Selberg} we find a quite compact result:
\begin{equation}\label{I1 computed}
\begin{split}
\textbf{I}_1&=(-)^{n-1}\frac{S_n(-\tfrac12,d-\tfrac52)}{S_n(-\tfrac12,d-\tfrac32)}\,\frac{n!(n+d-2)}{{\displaystyle \prod_k}(k+n+d-3)}\,P_{n-1}^{(1/2,d-3/2)}(1-2t)\rvert_{t=1}\\[2mm]
&=\frac{2n(n+d-2)}{2d-3}\;.
\end{split}
\end{equation}

We now turn to compute the average $\textbf{I}_2$, defined as
\begin{equation}\label{I2 defined}
\textbf{I}_2:=\left\langle\!\!\!\left\langle\sum_{i\neq j}\frac{1}{(1-x_i)(1-x_j)}\right\rangle\!\!\!\right\rangle_E =\textbf{J}-\textbf{K}\;.
\end{equation}
From the definition of $S_{n,1}(\alpha, \beta;t)$ in \eqref{Sn1}, it is easy to see that $\textbf{I}_2$ is related to its second $t$ derivative as
\begin{equation}\label{I2}
\textbf{I}_2=\frac{\mathcal{N}}{a_0}\,(-)^n\,\de^2_tS_{n,1}(-\tfrac12,d-\tfrac52;t)\rvert_{t=1}
=(-)^n\frac{\de^2_tS_{n,1}(-\tfrac12,d-\tfrac52;t)}{S_{n}(-\tfrac12,d-\tfrac32)}\;,
\end{equation}
and in the same way we computed $\textbf{I}_1$ we find for $\textbf{I}_2$
\begin{equation}\label{I2 computed}
\begin{split}
\textbf{I}_2&=\frac{S_n(-\tfrac12,d-\tfrac52)}{S_n(-\tfrac12,d-\tfrac32)}\,\frac{n!(n+d-2)(n+d-1)}{{\displaystyle \prod_k}(k+n+d-3)}\,P_{n-2}^{(3/2,d-1/2)}(1-2t)\rvert_{t=1}\\[2mm]
&=4n(n-1)\frac{(n+d-1)(n+d-2)}{(2d-1)(2d-3)}\;.
\end{split}
\end{equation}

We need at this point to introduce one further generalization of Selberg's integral, provided by Kaneko \cite{kaneko}:
\begin{equation}\label{Kn}
\begin{split}
K_n(\alpha,\beta;t)&:= \prod_{i=1}^n\int_0^1dx_i\,x_i^{\alpha}(1-x_i)^{\beta}(1-tx_i)^{-1}\prod_{k<l}(x_k-x_l)^2\\
&=S_n(\alpha,\beta)\,{}_2F_1(n,n+\alpha;2n+\alpha+\beta;t)\;,
\end{split}
\end{equation}
where ${}_2F_1(a,b;c;t)$ is the Gauss hypergeometric function. By taking two derivatives with respect to $t$ in \eqref{Kn} and evaluating at $t=1$ one finds an average that is related to $\textbf{K}$ by linear combinations of $\textbf{I}_1$ and $\textbf{I}_2$. We shall then define $\textbf{I}_3$ as
\begin{equation}\label{I3 defined}
\textbf{I}_3:=\frac{\mathcal{N}}{a_0}\,\de^2_tK_n(-\tfrac12,d-\tfrac12;t)\rvert_{t=1}=\frac{S_n(-\tfrac12,d-\tfrac12)}{S_n(-\tfrac12,d-\tfrac32)}\,\de^2_t
{}_2F_1(n,n-\tfrac12;2n+d-1;t)\rvert_{t=1}\;.
\end{equation}
In order to perform the computation we need the following properties of the hypergeometric function:
\begin{equation}\label{hyper func}
\begin{split}
\frac{d^k}{dz^k}\,{}_2F_1(a,b;c;z) &= \frac{(a)_k(b)_k}{(c)_k}\,{}_2F_1(a+k,b+k;c+k;z)\;,\quad (a_k):=\frac{\Gamma(a+k)}{\Gamma(a)}\;,\\[2mm]
{}_2F_1(a,b;c;1)&=\frac{\Gamma(c)\Gamma(c-a-b)}{\Gamma(c-a)\Gamma(c-b)}\;.
\end{split}
\end{equation}
Using now \eqref{hyper func} in \eqref{I3 defined}, we can compute $\textbf{I}_3$ that results
\begin{equation}\label{I3 computed}
\begin{split}
\textbf{I}_3&=\frac{S_n(-\tfrac12,d-\tfrac12)}{S_n(-\tfrac12,d-\tfrac32)}\,\frac{(n)_2(n-\tfrac12)_2}{(2n+d-1)_2}\,{}_2F_1(n+2,n+\tfrac32;2n+d+1;t)\rvert_{t=1}\\[2mm]
&=\frac{n(n+1)(4n^2-1)}{(2d-3)(2d-5)}\;.
\end{split}
\end{equation}
By using the definition \eqref{Kn} and taking the double derivative with respect to $t$ in $t=1$ one finds that $\textbf{K}$ is given as the following linear combination:
\begin{equation}\label{K relation}
\textbf{K}=\frac12\,\textbf{I}_3-\frac12\,\textbf{I}_2+(n+1)\,\textbf{I}_1-\frac{n(n+1)}{2}\;,
\end{equation}
whereas, by means of $\textbf{I}_2=\textbf{J}-\textbf{K}$, one has
\begin{equation}\label{J relation}
\textbf{J}=\frac12\,\textbf{I}_3+\frac12\,\textbf{I}_2+(n+1)\,\textbf{I}_1-\frac{n(n+1)}{2}\;.
\end{equation}
This concludes our computations of the modular integrals for even $N=2n$. Although the SDW coefficients can be read off straightforwardly from $\textbf{I}_1$, $\textbf{J}$ and $\textbf{K}$, we choose to present them in the paper in terms of $\textbf{I}_1$, $\textbf{I}_2$ and $\textbf{I}_3$, since they have much more compact expressions.

\subsection{Odd {\it N}}

We turn now to compute the modular integrals required for odd $N=2n+1$. The averages needed will have exactly the same structure as the even $N$ case, the only difference being the form of the modular measure. In particular, the only changes needed will be in the prefactor $\mathcal{N}$ and in all the generalized Selberg's formulas, where the parameter $\alpha$ will switch everywhere from $-\frac12$ to $+\frac12$. The averages in the odd case are explicitly given by
\begin{equation}\label{modular average x var odd}
\left\langle\!\left\langle f(x_j)\right\rangle\!\right\rangle_O:=\frac{\mathcal{N}}{a_0}\prod_{i=1}^n\int_0^1dx_i\,x_i^{1/2}(1-x_i)^{d-3/2}
\prod_{k<l}(x_k-x_l)^2f(x_j)\;,
\end{equation}
where we see that the only difference between \eqref{modular average x var odd} and \eqref{modular average x var} is the power $1/2$ instead of $-1/2$, that is the $\alpha$ parameter we used in all the previous computations. In addition, the prefactor now reads
\begin{equation}\label{Norm odd N}
\mathcal{N}=\frac{2^{2(d-1)+n(2n+2d-3)}}{\pi^nn!}\;.
\end{equation}

The averages have the same definition as before, being
\begin{equation}\label{angular averages odd}
\begin{split}
\widetilde{\textbf{I}}_1 &:= \left\langle\!\!\!\left\langle\sum_{i=1}^n\cos^{-2}\frac{\theta_i}{2}\right\rangle\!\!\!\right\rangle_O=
\left\langle\!\!\!\left\langle\sum_{i=1}^n\frac{1}{1-x_i}\right\rangle\!\!\!\right\rangle_O\;,\\[2mm]
\widetilde{\textbf{J}} &:= \left\langle\!\!\!\left\langle\sum_{i,j=1}^n\cos^{-2}\frac{\theta_i}{2}\cos^{-2}\frac{\theta_j}{2}\right\rangle\!\!\!\right\rangle_O=
\left\langle\!\!\!\left\langle\sum_{i,j=1}^n\frac{1}{(1-x_i)(1-x_j)}\right\rangle\!\!\!\right\rangle_O\;,\\[2mm]
\widetilde{\textbf{K}} &:= \left\langle\!\!\!\left\langle\sum_{i=1}^n\cos^{-4}\frac{\theta_i}{2}\right\rangle\!\!\!\right\rangle_O=
\left\langle\!\!\!\left\langle\sum_{i=1}^n\frac{1}{(1-x_i)^2}\right\rangle\!\!\!\right\rangle_O\;.
\end{split}
\end{equation}
Again, we will compute $\widetilde{\textbf{I}}_2$ and $\widetilde{\textbf{I}}_3$ instead of $\widetilde{\textbf{J}}$ and $\widetilde{\textbf{K}}$. The degrees of freedom factor $a_0$ now reads
\begin{equation}\label{a0 x var odd}
a_0=\mathcal{N}\prod_{i=1}^n\int_0^1dx_i\,x_i^{1/2}(1-x_i)^{d-3/2}\prod_{k<l}(x_k-x_l)^2\;.
\end{equation}
Everything goes in the same way as it did with even $N$, and we easily obtain:
\begin{equation}\label{all odd integrals}
\begin{split}
a_0&=\mathcal{N}S_n(\tfrac12,d-\tfrac32)=\frac{2^{d-2+n}}{d}\frac{(2d-2)!}{[(d-1)!]^2}\prod_{k=1}^{n-1}\frac{(k+d-1)(2k+1)!\,(2k+2d-3)!}{(2k+d-1)!\,(2k+d)!}\;,\\[2mm]
\widetilde{\textbf{I}}_1&=\frac{\mathcal{N}}{a_0}\,(-)^n\,\de_tS_{n,1}(\tfrac12, d-\tfrac52;t)\rvert_{t=1}=\frac{2n(n+d-1)}{2d-3}\;,\\[2mm]
\widetilde{\textbf{I}}_2&=\frac{\mathcal{N}}{a_0}\,(-)^n\,\de^2_tS_{n,1}(\tfrac12,d-\tfrac52;t)\rvert_{t=1}=\frac{4n(n-1)(n+d)(n+d-1)}{(2d-1)(2d-3)}\;,\\[2mm]
\widetilde{\textbf{I}}_3&=\frac{\mathcal{N}}{a_0}\,\de^2_tK_n(\tfrac12,d-\tfrac12;t)\rvert_{t=1}=\frac{n(n+1)(2n+1)(2n+3)}{(2d-3)(2d-5)}\;.
\end{split}
\end{equation}
Also the relations that give $\widetilde{\textbf{J}}$ and $\widetilde{\textbf{K}}$ remain unchanged and are
\begin{equation}\label{K relation odd}
\begin{split}
\widetilde{\textbf{K}}&=\frac12\,\widetilde{\textbf{I}}_3-\frac12\,\widetilde{\textbf{I}}_2+(n+1)\,\widetilde{\textbf{I}}_1-\frac{n(n+1)}{2}\;,\\[2mm]
\widetilde{\textbf{J}}&=\frac12\,\widetilde{\textbf{I}}_3+\frac12\,\widetilde{\textbf{I}}_2+(n+1)\,\widetilde{\textbf{I}}_1-\frac{n(n+1)}{2}\;.
\end{split}
\end{equation}


\end{document}